# Contribution of the oxygen reduction reaction to the electrochemical cathodic partial reaction for Mg-Al-Ca solid solutions

Markus Felten*, Siyuan Zhang**, Rasa Changizi**, Christina Scheu**, Mark Bruns***, Michael Strebl***, Sannakaisa Virtanen***, Daniela Zander*°

*Chair of Corrosion and Corrosion Protection, Division of Materials Science and Engineering, Foundry Institute Aachen, RWTH Aachen University, Aachen Germany

**Max-Planck-Institut für Eisenforschung, Düsseldorf Germany

*** Department of Materials Science, Institute for Surface Science and Corrosion, Friedrich-Alexander-Universität Erlangen-Nürnberg, Erlangen Germany

° Correspondence: d.zander@gi.rwth-aachen.de

**Abstract**

The electrochemical corrosion rate of Magnesium (Mg) and Mg alloys depends on the stability of the formed surface layer. Based on the Mg substrate, the oxide structure comprises a dense $MgO/Mg(OH)_2$ mixture underneath a porous plate-like $Mg(OH)_2$ layer. While the kinetics of the anodic partial reaction has been mainly attributed to MgO, recent studies showed an effect of the $Mg(OH)_2$ layer thickness on the cathodic partial reaction. A thinner $Mg(OH)_2$ layer has been associated with a higher kinetics of the oxygen reduction rate. In the present study, the proposed mechanism has been further investigated via in situ respirometric measurements with Mg-Al-Ca solid solution in electrolytes with different pH values (pH = 8-13). The results indicate an additional effect based on the structure of the surface layer in the passive state of Mg corrosion. Furthermore, two different Al enriched interlayers at the Mg/MgO- and $MgO/Mg(OH)_2$ interfaces were observed and discussed in terms of their thermodynamic stability under alkaline immersion conditions.

Magnesium and Mg alloys represent a potential versatile material class for lightweight applications [1-3]. Nevertheless, the high chemical activity and the merely partially protective surface film still restrict the widespread application of Mg alloys [4-6]. The surface film formed under neutral immersion conditions has been either described as a three-layered film composed of an hydrated inner cellular $Mg(OH)_2$ layer, a middle MgO layer and an outer plate-like $Mg(OH)_2$ layer [7] or as an bi-layered film comprising an inner nanocrystalline $MgO/Mg(OH)_2$ mixture beneath an outer plate-like $Mg(OH)_2$ layer [8]. The latter layer precipitated from the electrolyte after exceeding the solubility limit of $Mg^{2+}$ cations and hydroxide ions within the electrolyte [7]. Higher pH values increase the corrosion resistance due to a more stable passive film [6, 9, 10], while $Mg(OH)_2$ is thermodynamically stable at pH > 10.5 and forms a quasi-passive layer, according to Pourbaix diagram [11]. The cathodic partial reaction has been generally stated to be dominated by the hydrogen evolution reaction (HER) [12-14]. However, recent studies observed a significant oxygen reduction reaction (ORR) contribution to the total cathodic partial reaction up to 29.1 % for ultra-high purity (UHP) Mg in 0.05 M NaCl (pH = 5.8) [15]. The proposed underlying mechanism assumes a restricted oxygen diffusion from the bulk electrolyte to the metal/electrolyte interface as a consequence of the outer plate-like $Mg(OH)_2$ layer formation in commercially pure (CP) Mg. UHP-Mg, however, exhibits a lower corrosion rate and therefore exhibits merely a ~70 nm $Mg(OH)_2$ layer thickness [16], which subsequently lowers the oxygen diffusion restriction and causes the higher ORR contribution.

For Mg-Al alloys instead, a further Al-enriched Mg/MgO interface has been observed in the native oxide layer [17-19] and under neutral immersion conditions [17], which tends to increase the electrochemical corrosion resistance. Song et al. [20] observed an Al-enriched Mg/MgO interface also under alkaline immersion conditions, where Al is thermodynamically unstable according to the classic Pourbaix diagram [21]. The authors assume that the Al layer is not in direct contact with the alkaline solution and therefore preserves the integrity [20]. Recent DFT-based surface phase diagram calculations [22] however, describe an energy gain upon solvation for Al-substituted Mg surfaces, which promotes the Al segregation to the top surface. Under consideration of Al species in the electrolyte, these calculations predict an Al monolayer surface structure as the thermodynamically most stable one and therefore predict an Al enrichment at the surface specifically because the surface is in contact with the electrolyte.

In this study, the ORR/HER contribution to the total cathodic partial reaction has been investigated for a Mg-Al-Ca (CP) solid solution in electrolytes of pH = 8.0 – 13.0. Thereafter, we critically review the previously proposed underlying mechanism [16] for an enhanced ORR contribution as a consequence of a reduced corrosion current density for Mg corrosion under quasi-passive immersion conditions (pH > 10.5). Furthermore, the principle of the Al accumulation at the Mg/MgO interface has been examined, to provide experimental evidence for both proposed mechanisms describing the stability under alkaline immersion conditions [20, 22].

The Mg-2Al-0.1Ca (CP) alloy in Table 1 has been investigated. The alloy was manufactured via inductive melting under Argon (Ar) atmosphere, heat treated at 450 °C for 20 min, hot rolled to a total deformation of 50 %, further homogenized at 500 °C under Ar atmosphere for 24 h and ultimately water quenched. All specimens were metallographically prepared final polished according to ref. [23] and stored under moderate atmospheric condition for ~24 h, to establish stable native oxide surface conditions as recommended in refs. [19, 24]. The ORR/HER contribution under immersion conditions at the free corrosion potential was investigated via in situ respirometric measurements for 6 h in the electrolytes presented in Table 2. Respirometric measurements reliably monitor the corrosion rate by differentiating the cathodic partial reaction in HER and ORR [25-27]. The oxygen partial pressure was measured by an optical $O_2$ meter (Piccolo2OEM, Pyroscience, Germany) and the $H_2$ pressure was derived from the total pressure with a digital pressure sensor (Bosch Sensortec, BMP280, Germany) in a closed system. The molar changes associated with the consumed $O_2$ and evolved $H_2$ are determined by the ideal gas law and subsequently transformed to charges via Faraday's law, while the dissolved hydrogen is estimated by Henry's law. The diffusion of oxygen from the gas phase into the electrolyte has been furthermore monitored without a specimen in the closed system and revealed an equilibrium after $t_{eq}$= 4 h with a constant oxygen concentration of $c_{O2} = 262 \pm 2$ µmol/l, independent of the utilized electrolyte. Thus, $t_{eq}$ represents the immersion time when the ORR values are reliably determined. To further examine the quasi-passive layer formed under alkaline immersion conditions, the surfaces after an immersion time of 6 h in the electrolytes with pH = 11.5 and pH = 13 (V = 200 ml) were analyzed. Scanning transmission electron microscopy (STEM) was performed using a probe-corrected Titan Themis microscope operated at 300 kV. Cross-section samples were prepared using a Scios2 focused ion beam, after ~1 µm carbon marker was deposited to protect the specimen surface [28]. High angle annular dark-field (HAADF)-STEM images were acquired using a detector range of 73-200 mrad. STEM-energy dispersive X-ray (EDX) spectrum imaging was recorded using the SuperX-detector. Multivariate statistical analysis [29] was applied to denoise the dataset and highlight the Al-enriched layers.

Table 1: Chemical composition and impurity concentration determined via inductively coupled plasma (ICP)-mass spectrometry (MS) and ICP-optical emission spectrometry (OES).

|  | Mg [wt.%] | Al [wt.%] | Ca [wt.%] | Fe [ppm] | Cu [ppm] | Mn [ppm] | Ni [ppm] | Si [ppm] |
|---|---|---|---|---|---|---|---|---|
| Mg-2Al-0.1Ca | Bal. | 2.14 | 0.11 | 68 | 5 | 63 | < 5 | 59 |

Table 2: Utilized electrolytes and their chemical composition.

|   | Electrolyte | Components | pH [-] | Volume [ml] |
|---|---|---|---|---|
| 1 | Borate Buffer | $H_3BO_3$ (52 mmol/l) $Na_2B_4O_7$ (1.5 mmol/l) | 8.0 ± 0.1 | 80 |
| 2 | Borate Buffer + KOH | $H_3BO_3$ (52 mmol/l) $Na_2B_4O_7$ (1.5 mmol/l) KOH (66 mmol/l) | 9.5 ± 0.1 | 80 |
| 3 | KOH | KOH (3 mmol/l) | 11.5 ± 0.1 | 0.02 |
| 4 | KOH | KOH (100 mmol/l) | 13.0 ± 0.1 | 0.02 |

Figure 1 exhibits the in situ respirometric results as a function of the immersion time and the pH. The determined HER/ORR charges were converted to the corresponding current densities in the time interval of 5-6 h and presented in Table 3. While a general decrease of the total current density ($i_{tot}$) with increasing pH has been observed, a pronounced $i_{tot}$ drop occurs at the active-passive transition between pH =9.5 and pH =11.5. Furthermore, with an increasing pH and subsequently reduced $i_{tot}$, the relative ORR fraction increases from 1.9 % to 38.7 %, which is in line with the observation in [16]. However, the absolute $i_{ORR}$ does not increase with decreasing $i_{tot}$, as expected by the proposed ORR mechanism [15, 16]. In contrast, the $i_{ORR}$ in the passive state is significantly reduced by a factor of ~30 compared to the active state, while no significant difference between pH = 11.5 and pH = 13.0 is observed. Effects on the ORR kinetics associated with the low electrolyte volumes (Table 2) are excluded, as these would exclusively require consideration for very thin electrolyte layers comprising merely a few monolayers [30]. Further interferences with varying oxygen uptake reactions at the electrolyte/gas phase interface are as well excluded, as these effects restrict the limiting current density in higher orders of magnitudes in the range of $i_{ORR} = 2.5$ mA/cm² [30]. Furthermore, as all utilized electrolytes exhibit the same equilibrated oxygen concentration, the same driving force for the diffusion-based mass transportation is expected for $t_{eq} > 4$ h, so that exclusively material-inherent properties cause the observed $i_{ORR}$ results.

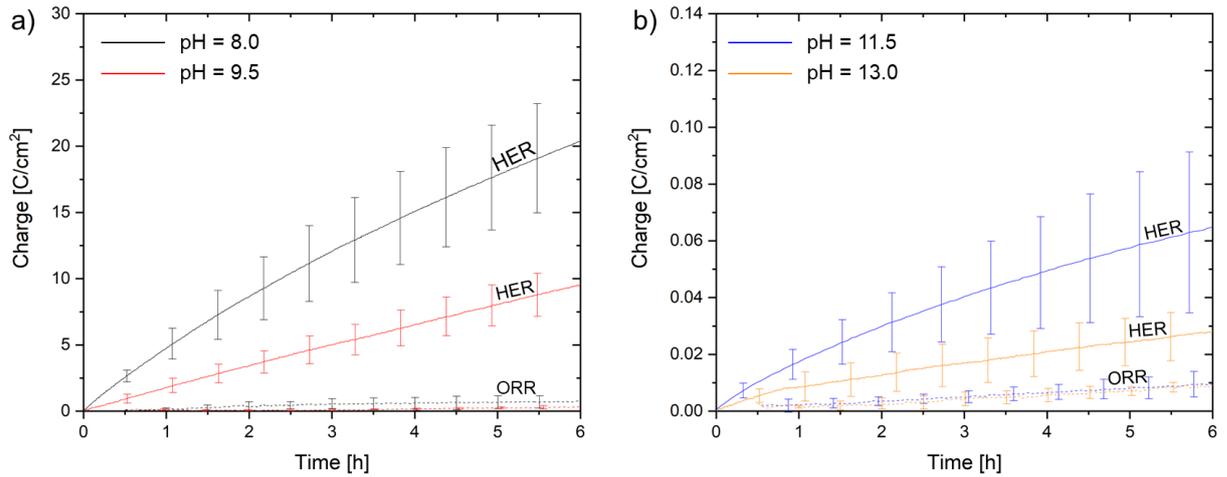

Figure 1 In situ respirometry results for Mg-Al-Ca solid solution as a function of immersion time in a) pH=8.0 / 9.5 and b) pH=11.5 / 13.0.

Table 3 HER and ORR current density (i) evaluated in the time interval of 5-6 h for Mg-Al-Ca solid solution.

| pH [-] | $i_{tot}$ [μA/cm$^2$] | $i_{HER}$ [μA/cm$^2$] | $i_{ORR}$ [μA/cm$^2$] | $i_{ORR}/(i_{ORR}+i_{HER})$ [%] |
|---|---|---|---|---|
| 8.0 | 726 ± 104 | 712 ± 91 | 13.7 ± 13.6 | 1.9 |
| 9.5 | 434 ± 73 | 411 ± 58 | 23.9 ± 15.0 | 5.5 |
| 11.5 | 2.4 ± 1.7 | 2.0 ± 1.3 | 0.4 ± 0.38 | 16.5 |
| 13.0 | 1.7 ± 0.5 | 1.0 ± 0.5 | 0.7 ± 0.02 | 38.7 |

Figure 2 exhibits the quasi-passive layer formed under alkaline immersion conditions. The quasi-passive layer observed in pH = 11.5 (Figure 2 a)) reveals a ~50 nm thick Mg(OH)$_2$ layer formed of ~400 nm long Mg(OH)$_2$ flakes, precipitating from the electrolyte [7]. While the initial Mg(OH)$_2$ flakes sink horizontal on the specimen surface, the subsequently formed Mg(OH)$_2$ flakes vertically line up, forming the well-known porous Mg(OH)$_2$ structure [7, 8]. On the other hand, the quasi-passive layer formed in pH =13 (Figure 2 b)) exhibits much shorter Mg(OH)$_2$ flakes on the surface, demonstrating the reduced tendency for a Mg(OH)$_2$ precipitation with decreasing corrosion current density. As $i_{ORR}$ is equivalent for both alkaline immersion conditions and $i_{ORR}$ is lower compared to the active state, where a thicker Mg(OH)$_2$ flake layer is expected, this layer does not exclusively cause the determined reduced $i_{ORR}$ under alkaline immersion conditions, as proposed for CP-Mg under neutral immersion conditions [16].

The dense MgO/Mg(OH)$_2$ layer in Figure 2 comprises a similar thickness of ~50 nm in both investigated alkaline electrolytes, similar to the thickness determined in ref. [16]. However, the EDX measurements in Figure 2 exhibit O ~ 66 at.%, indicating a dense and pure Mg(OH)$_2$ layer as expected at pH > 10.5, according to Pourbaix diagram [11]. As the flake like Mg(OH)$_2$ layer is supposed to decrease the O diffusion kinetic under neutral immersion conditions [16], it appears likely that the dense Mg(OH)$_2$ layer formed under alkaline immersion conditions, further decreases the O diffusion and therefore cause the observed reduced $i_{ORR}$ under these conditions. A further explanation might as well account for the low $i_{ORR}$ under alkaline immersion conditions. A higher free electrode potential for Mg-Al-Ca alloys is highly probable under passive immersion conditions compared to active dissolution conditions, which would consequently reduce the overpotential of the ORR reaction and might therefore cause the determined lower absolute $i_{ORR}$ in the quasi-passive state of Mg corrosion. To distinguish between both proposed mechanisms, future studies will combine in situ respirometric experiments with a standard three-electrode setup to adjust the electrode potential and simultaneously monitor $i_{ORR}$ under alkaline immersion conditions.

Both quasi-passive layers in Figure 2 formed under alkaline immersion conditions exhibit two discontinuous and ~20 nm thick Al enriched layers at the Mg/MgO- and MgO/Mg(OH)$_2$ interface, respectively. While the first Al-enriched interface was previously observed under alkaline immersion conditions by Song et al. [20], the latter Al-enriched interface has not yet been reported. As the Al layer at the MgO/Mg(OH)$_2$ interface exists at the former native oxide interface, it most likely represents the Al enrichment formed under atmospheric conditions [17-19]. However, as this surface is certainly in contact with the electrolyte, these results contradict the assumption that the classically thermodynamic unstable Al-enriched interface merely preserves its integrity as it is not in direct contact with the alkaline solution [20]. Instead, there seems to be a rather different driving force for Al at the MgO interfaces. The DFT based surface phase diagram calculations in ref. [22] predict an Al monolayer surface structure under electrochemical corrosion conditions in the Mg crystal, which might account for the Al enrichment at the Mg/MgO interface. However, instead of a monolayer in the Mg crystal, a selectively interrupted ~20 nm thick enrichment has been observed. The difference to the DFT calculation might arise as a consequence of the adjacent MgO layer, which potentially enables a lateral diffusion of the Al to specific discontinuous domains triggered by induced lattice mismatches. DFT calculations for the observed Al enrichment at the MgO/Mg(OH)$_2$ interface are currently not available, but might reveal a similar tendency for an Al accumulation in a MgO crystal.

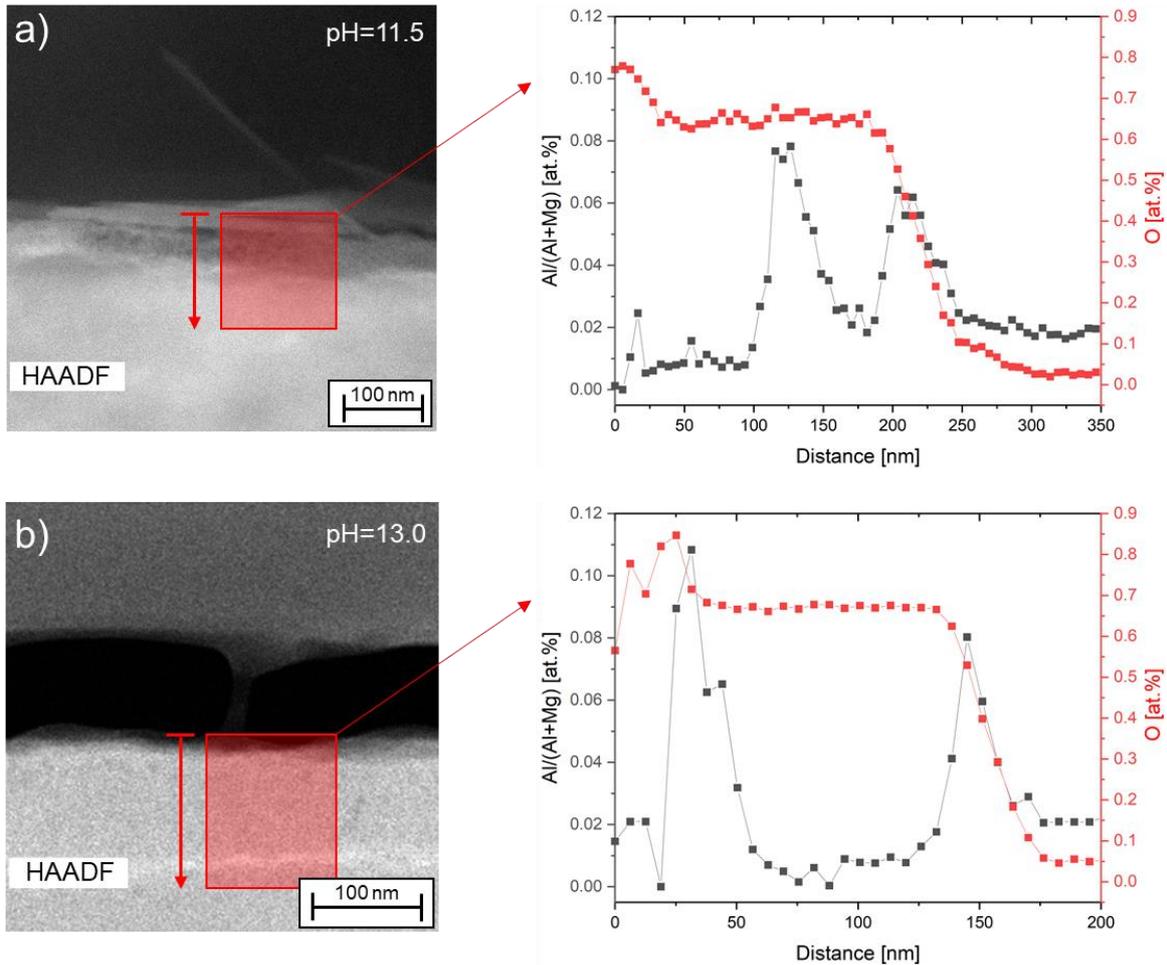

Figure 2 TEM analysis of Mg-Al-Ca solid solution after 6 h immersion in a) pH = 11.5 and b) pH = 13.0.

In summary, the ORR/HER contribution to the cathodic partial reaction was successfully investigated for a Mg-Al-Ca (CP) solid solution in electrolytes with different pH values (pH = 8.0-13.0). The previously proposed mechanism of a reduced ORR contribution with a thicker $Mg(OH)_2$ flake layer, as a consequence of a decreased corrosion current density [16], appears not exclusively valid in the passive state of Mg corrosion. While the relative ORR contribution increases with a reduced corrosion current density, the absolute $i_{ORR}$ current density decreases in the passive state. This effect can be attributed to the dense $Mg(OH)_2$ oxide layer or to a reduced overpotential of the ORR reaction under alkaline immersion conditions. Furthermore, an Al-enriched Mg/MgO interface was observed under alkaline immersion conditions as reported in ref. [20]. However, a second Al enrichment was determined at the $MgO/Mg(OH)_2$ interface, which might originally stem from the native oxide. As the Al enrichment apparently does not directly dissolve, when in contact with the alkaline electrolyte, it rather exhibits a thermodynamic stable state at the $MgO/Mg(OH)_2$-, Mg/MgO interfaces as calculated for the Mg bulk.

Acknowledgements

The authors gratefully acknowledge the financial support of the Deutsche Forschungsgemeinschaft (DFG) within projects B01, B05 and C03 of the Collaborative Research Center (SFB) 1394 "Structural and Chemical Atomic Complexity - from defect phase diagrams to material properties" – project ID 409476157. Furthermore, the authors acknowledge the DFG for the research training group GRK 1896 "In situ microscopy with electrons, X-rays and scanning probes" - project-ID 218975129. Moreover, the authors thank Hauke Springer from the Institute of Metal Forming at RWTH Aachen University for providing the investigated material.